\documentclass[letterpaper, 10 pt, conference]{ieeeconf}  

\IEEEoverridecommandlockouts                              
\usepackage{amsmath} 
\usepackage{amssymb}  
\usepackage{mathptmx} 
\usepackage{subfigure}
\usepackage{mathrsfs} 
\usepackage{float} 
\usepackage{multirow} 
\usepackage{graphicx}
\usepackage{microtype} 
\usepackage{algorithm}
\usepackage{algpseudocode}

\algrenewcommand\alglinenumber[1]{#1:}
\usepackage{color}

\newtheorem{remark}{Remark}

\newcounter{rmnum}
\newenvironment{romannum}{\begin{list}{{\upshape (\roman{rmnum})}}{\usecounter{rmnum}
			\setlength{\leftmargin}{14pt}
			\setlength{\rightmargin}{8pt}
			\setlength{\itemsep}{2pt}
			\setlength{\itemindent}{-1pt}
	}}{\end{list}}

\newcounter{anum}

\newcommand{\ud}{\,\mathrm{d}}

\def\v{{\sf K}}

\def\v{{\sf K}}

\def\Re{\mathbb{R}}

\def\Expect{{\sf E}}

\def\clZ{{\cal Z}}

\newcommand{\dt}[1]{\frac{\ud{1}}{\ud t}}

\newcommand{\HminN}{\underbar{H}^{(N)}}
\newcommand{\JN}{J^{(N)}}
\newcommand{\HN}{H^{(N)}}
\newcommand{\thetaN}{\theta^{(N)}}
\newcommand{\cN}{c^{(N)}}
\newcommand{\hatHN}{\hat{H}^{(N)}}
\newcommand{\hatHminN}{\hat{\underbar{H}}^{(N)}}

\title{\LARGE \bf
	Q-learning for POMDP: An application to learning locomotion gaits
}

\author{Tixian Wang, Amirhossein Taghvaei, and Prashant G. Mehta
\thanks{Financial support from the NSF grant CMMI-1462773 and ARO
  grant W911NF1810334 is gratefully acknowledged.}
\thanks{T.~Wang, A.~Taghvaei and P.~G.~Mehta are with the 
    Coordinated Science Laboratory and the Department of 
    Mechanical Science and Engineering at the University of 
    Illinois at Urbana-Champaign (UIUC).
	{\tt\small tixianw2@illinois.edu; taghvae2@illinois.edu; mehtapg@illinois.edu}}
}

\begin{document}
	
\maketitle
\thispagestyle{empty}
\pagestyle{empty}

\begin{abstract}

This paper presents a Q-learning framework for learning 
optimal locomotion gaits in robotic systems modeled as
coupled rigid bodies. Inspired by prevalence of periodic
gaits in bio-locomotion, an open loop periodic input is assumed to (say)
affect a nominal gait. The learning problem is to learn a
new (modified) gait by using only partial noisy measurements of the
state. The objective of learning is to maximize a given reward
modeled as an objective function in optimal control
settings. The proposed control architecture has three main
components: (i) Phase modeling of dynamics by a single
phase variable; (ii) A coupled oscillator feedback particle
filter to represent the posterior distribution of the phase
conditioned in the sensory measurements; and (iii) A
Q-learning algorithm to learn the approximate optimal
control law. The architecture is illustrated with the aid
of a planar two-body system. The performance of the
learning is demonstrated in a simulation environment.


    
\end{abstract}

\section{INTRODUCTION}

Biological locomotion is the movement of an animal from one location
to another location, through periodic changes in the shape of the
body, along with interaction with the environment~\cite{holmes2006dynamics}. 
The periodic motion of the shape, which constitutes the building block of 
locomotion, is called the {\it locomotion gait}. 
Examples of the locomotion gait are legged locomotion, flapping of the
wings for flying, or wavelike motion of the fish for swimming.  
It is a wonderful example for learning because the dynamics are
complicated but the goals (reward function) are easily modeled.  In
direct application of Q-learning to these problems, however, a problem
arises because the full state is not available.


The dynamics for such robotic systems are modeled using coupled rigid bodies.  
The configuration space is split into two sets of variables: 
(i) the {\it shape} variable that describes the robot's internal degrees 
of freedom; 
(ii) and the {\it group} variable that describes the global position and 
orientation of the robot. 
The dynamics of the shape variable is given by a second-order differential 
equation driven by control inputs. The dynamics of the group variable is 
given by a first-order differential equation governed by non-holonomic 
constraints in the system (e.g conversation laws or no slipping conditions).  

A typical approach to design the locomotion gait is to search for a periodic 
orbit in the shape space that leads to a desired {\it net} 
change in the group variable. This idea of producing a net change through 
underlying periodic motion is known as {\it mechanical rectification}
~\cite{brockett2003pattern,krishnaprasad1997motion} and the net change in 
the group variable is called {\it geometric phase}~\cite{kelly1995geometric,
krishnaprasad1990geometric,marsden1990reduction}. The optimal gait is
obtained by searching over a parameterized set of 
trajectories in the shape space to optimize a given optimization criteria ~\cite{blair2011optimal,murray1993nonholonomic,
walsh1995reorienting,melli2006motion,ostrowski1994nonholonomic,
hatton2010generating}. The resulting control law is open-loop and can
suffer from issues due to uncertainties in the environment, or
disturbance which perturbs the trajectory away from the orbit.  

The problem considered in this paper is to learn an approximate 
optimal gait given an open-loop periodic input. We do not
assume any control over this input: for example, it may correspond to
the nominal gait or it may be exerted from the environment.  The
presence of the periodic input creates a limit cycle in the
high-dimensional configuration space.  The control problem is to
actuate some of the system parameters to learn new types of gaits.  

For the purpose of learning, no knowledge of dynamic models is
assumed.  Furthermore, knowledge of full state is not assumed.  At
each time, one only has access to partial noisy measurements of the
state.  The proposed control architecture builds upon our prior work
on phase estimation~\cite{tilton2014control} and its use for optimal
control of bio-locomotion~\cite{taghvaei2014coupled}.  As
in~\cite{taghvaei2014coupled}, the control problem is modeled as an
optimal control problem.  Since full state feedback is not assumed, we
are in the partially observed settings.  The original
contribution of this work is to extend our prior work to a reinforcement
learning framework whereby new gaits can be learned by the robot only
through the use of noisy measurements and observed rewards.

The proposed control architecture has three parts:

\noindent
\textbf{1. Phase modeling:} Under the assumed periodic input, the
shape trajectory is a limit cycle in the high-dimensional phase
space.  The main complexity reduction technique is to introduce a
phase variable $\theta(t)\in [0,2\pi]$ to parametrize the
low-dimensional periodic orbit.  The inspiration comes from
neuroscience where phase reduction is a popular technique to obtain
reduced order model of neuronal
dynamics~\cite{izhikevich2007dynamical,brown2004phase}.


\medskip
\noindent
\textbf{2. Coupled oscillator feedback particle filter:} We construct
a nonlinear filtering algorithm to approximate the posterior
distribution of the phase variable $\theta(t)$, given time history of
sensory measurements.  The coupled oscillator feedback particle filter
(FPF) is comprised of a system of $N$ coupled oscillators
~\cite{tilton2013multi,tilton2012filtering}. The empirical distribution 
of the oscillators is used to approximate the posterior distribution. 

\medskip
\noindent
\textbf{3. Reinforcement learning:} The control problem is cast as a
partially observed optimal control problem.  The posterior
distribution represents an information state for the problem.  A
Q-learning algorithm is proposed where the Q-function is approximated
through a linear function approximation.  The weights are learned by
implementing a gradient descent algorithm to reduce the Bellman
error~\cite{watkins1989learning,vrabie2007continuous,mehta2009q}.  


The key component of the proposed architecture is the system of 
coupled oscillators that is used to both represent the posterior 
distribution (the belief state) and to learn the optimal control policy. 
This control system can be viewed as a central pattern generator (CPG)
which integrates sensory information to learn closed-loop optimal control 
policies for bio-locomotion.  

The proposed CPG control architecture is illustrated with the aid of 
a two-body planar system depicted in Fig.~\ref{fig:2-body}. 
An open-loop periodic torque is applied 
at the joint 
connecting the two links, which results in the body to oscillate in a 
periodic (but uncontrolled) manner. The control objective is to turn
the head clockwise by actuating the length of the tail body.  
Although only a two-body system is considered in this paper, the
proposed architecture can easily be generalized to coupled body models
of snake-robots and swimming fish robots. This is the subject of ongoing work.
The remainder of this paper is organized as follows: 
The problem is formulated in Section~\ref{sec:ProbForm}. 
The proposed solution is described in Section~\ref{sec:SolAppr}.
The numerical results appear in Section~\ref{sec:Numerics}. 

\section{Problem Formulation}
\label{sec:ProbForm}

\subsection{Modeling and dynamics}
\label{subsec:Model-Dyn}

\begin{figure}[t]
	\centering
	\includegraphics[width=1.0\columnwidth]{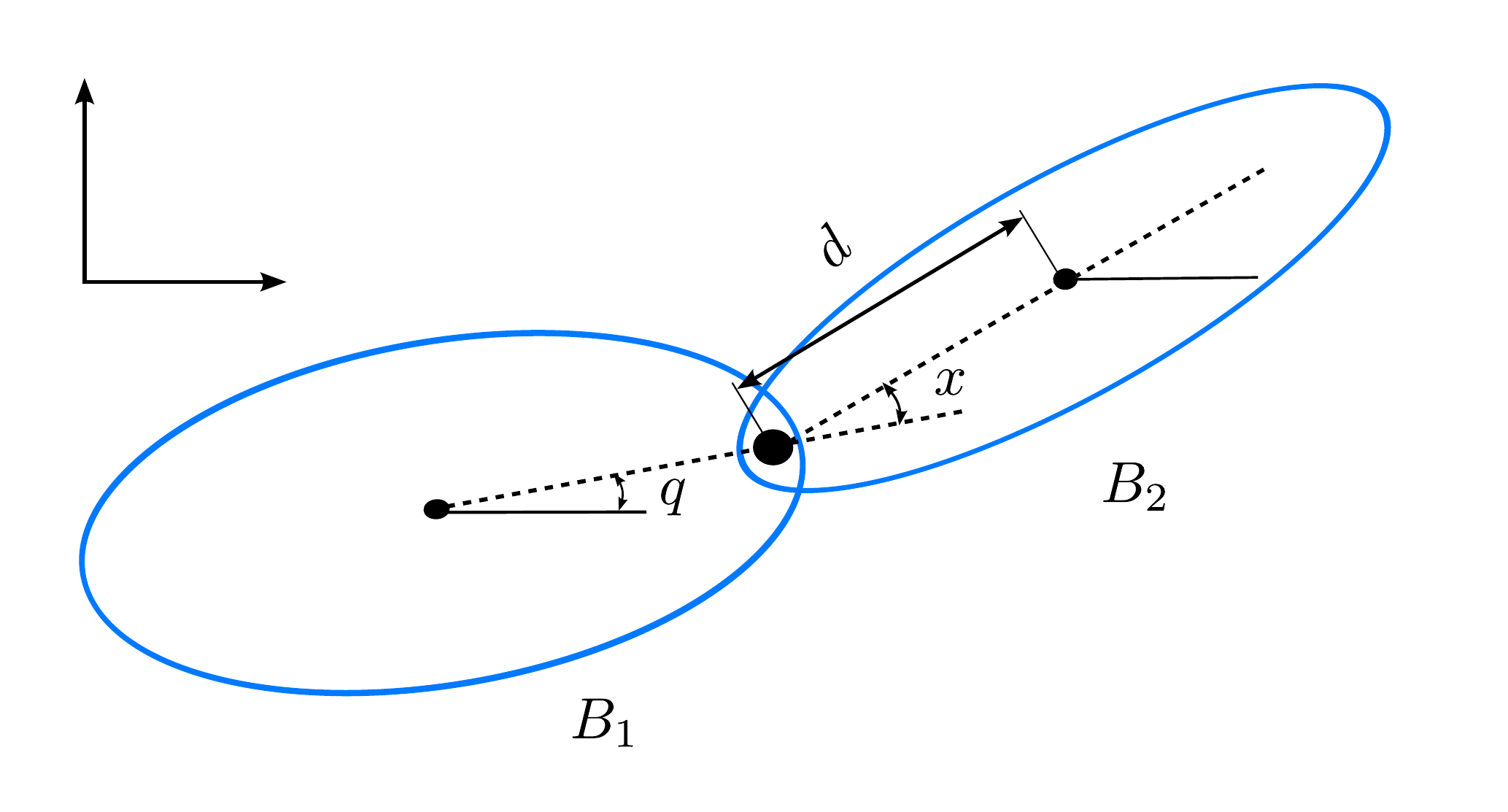}
	\caption{Schematic of the two-body system.}
	\label{fig:2-body}
\end{figure}

Consider a system of two planar rigid bodies, the head body $B_1$ and
the tail body $B_2$, connected by a single degree of freedom pin joint
as depicted in Figure~\ref{fig:2-body}. 
The configuration variables of the system are divided into two sets: 
i) the shape variable; ii) and the group variable.  For the two-body
system, the shape variable is the relative orientation between two
bodies.  It is denoted as the angle $x$. 
The group variable is the absolute orientation of the frame rigidly 
affixed to head body.  It is denoted as the angle $q$.

There is no external force applied to the system, which means that the 
total angular momentum is conserved. 
The joint is assumed to be actuated by a motor with driven torque $\tau$. 
An open-loop periodic input is assumed for the torque actuation:
\begin{equation}
    \tau(t) = \tau_0 \sin(\omega_0t)
    \label{eq:periodic_torque}
\end{equation}
where $\omega_0$ is the frequency and $\tau_0$ is the amplitude. 
There is no control objective associated with this torque input, 
except to have the shape variable $x$ oscillate in a periodic manner.

It is assumed that there exists a control actuation that changes the length 
of the tail body $B_2$ according to
\begin{equation}
    d(t) = (1+u(t))\bar{d}
    \label{eq:change_len}
\end{equation}
where $\bar{d}$ is the nominal length and $u$ represents the control input.

The dynamics of the two-body system is described by a second-order
ODE for the shape variable and a first-order ODE for the group variable:
\begin{subequations}
    \begin{align}
        \ddot{x}(t) & = \tilde{g}(x(t),\dot{x}(t),\tau(t),u(t)),\quad 
            (x(0),\dot{x}(0)) = (x_0,\dot{x}_0)
	    \label{eq:i-Dynamics} \\
    	\dot{q}(t) & = \tilde{f}(x(t),\dot{x}(t),u(t))
	    \label{eq:g-Dynamics}
	\end{align}
    \end{subequations}
The explicit form of the functions $\tilde{f}$ and $\tilde{g}$ and their 
derivations appear in~\cite{taghvaei2014coupled}. In this paper, the
explicit form of $\tilde{f}$ and $\tilde{g}$ are assumed unknown.
Rather, the dynamic model is treated as a simulator (black-box) that
is used to simulate the 
dynamics~\eqref{eq:i-Dynamics}-\eqref{eq:g-Dynamics}.


\medskip

\begin{remark}
The modeling procedure is easily generalized to a chain of $n$ planar rigid 
links, e.g used to model snake robots as in~\cite{kelly1995geometric}. In such 
systems, the group variable is the position and orientation of the robot, 
and the configuration space of shape variable is $n-1$ dimensional. 
With $n$ links, the dimension of the system is $2n+1$. An interesting problem 
for the snake robot is to learn a turning maneuver by changing the friction 
coefficients with respect to the surface.
\end{remark}

\subsection{Observation process}
\label{subsec:obsv}

For the purposes of learning and control design, the state
$(x,\dot{x})$ is assumed to be unknown.  The following
continuous-time observation model is assumed for the sensor: 
\begin{equation}
    \ud Z(t) = \tilde{h}(x(t),\dot{x}(t))\ud t + \sigma_W \ud W(t)
    \label{eq:obsv}
\end{equation}
where $Z(t)\in \Re$ denotes the observation at time $t$, $W(t)$ is a standard 
Wiener process, and $\sigma_W>0$ is the noise strength. 
The observation function $\tilde{h}:\Re^2\to\Re$ is known and is assumed to be
$C^1$. 

\subsection{Optimal control problem}
\label{subsec:CtrlProb}

The control objective is to turn the head body $B_1$ clockwise by
actuating the control input $u(t)$.  An uncontrolled periodic torque
input~\eqref{eq:periodic_torque} is assumed to be present.  The
control objective is modeled as a discounted infinite-horizon optimal
control problem:
\begin{equation}
    \tilde{J}(x_0,\dot{x}_0) 
    = \underset{u(\cdot)}{\min}\ {\sf E} \left[ 
    \int_0^\infty e^{-\gamma t} \tilde{c}(x(t),\dot{x}(t),u(t)) \ud t \right]
    \label{eq:opt-cont}
\end{equation}
subject to the dynamic constraints~\eqref{eq:i-Dynamics}.  Here, 
$\gamma>0$ is the discount rate and the cost function 
\[
\tilde{c}(x,\dot{x},u)=\tilde{f}(x,\dot{x},u)+\frac{1}{2\epsilon}u^2
\] 
with $\epsilon>0$ as the control penalty parameter. 
The minimum is over all control inputs $u(\cdot)$ adapted to the filtration 
$\mathcal{Z}_t:=\sigma(Z(s);s\in[0,t])$ generated by the observation process. 

The particular form of the cost function is assumed to maximize the
rate of change, in the clockwise direction, in the group variable
$q$.  Indeed, by~\eqref{eq:g-Dynamics},
$\tilde{f}(x,\dot{x},u)=\dot{q}$.  Therefore, minimizing the cost
leads to net negative change in the head orientation $q(t)$.  This
corresponds to the clockwise rotation.    

\section{Solution Approach}
\label{sec:SolAppr}
Solving the optimal control problem~\eqref{eq:opt-cont} is challenging
because of the following reasons:
\begin{enumerate}
	\item The function $\tilde{g}$ in the dynamic model~\eqref{eq:i-Dynamics} 
	is not known.  For coupled rigid bodies, the model is nonlinear
        and complicated due to the details of the geometry, models of
        contact forces with the environment, uncertain parameters,
        etc.  
	\item The problem is partially observed, i.e., the 
	state $(x,\dot{x})$ is not known.  As the number of links
        grow, the state can be very high dimensional.  
	\item The explicit form of the function $\tilde{f}$ which captures the relationship between shape and group dynamics is not known. 
          
\end{enumerate}

These challenges are tackled by the following three-step procedure:

\subsection{Step 1. Phase modeling}
\label{subsec:model-reduction}

\begin{figure}[t]
	\centering
	\includegraphics[width=0.6\columnwidth]{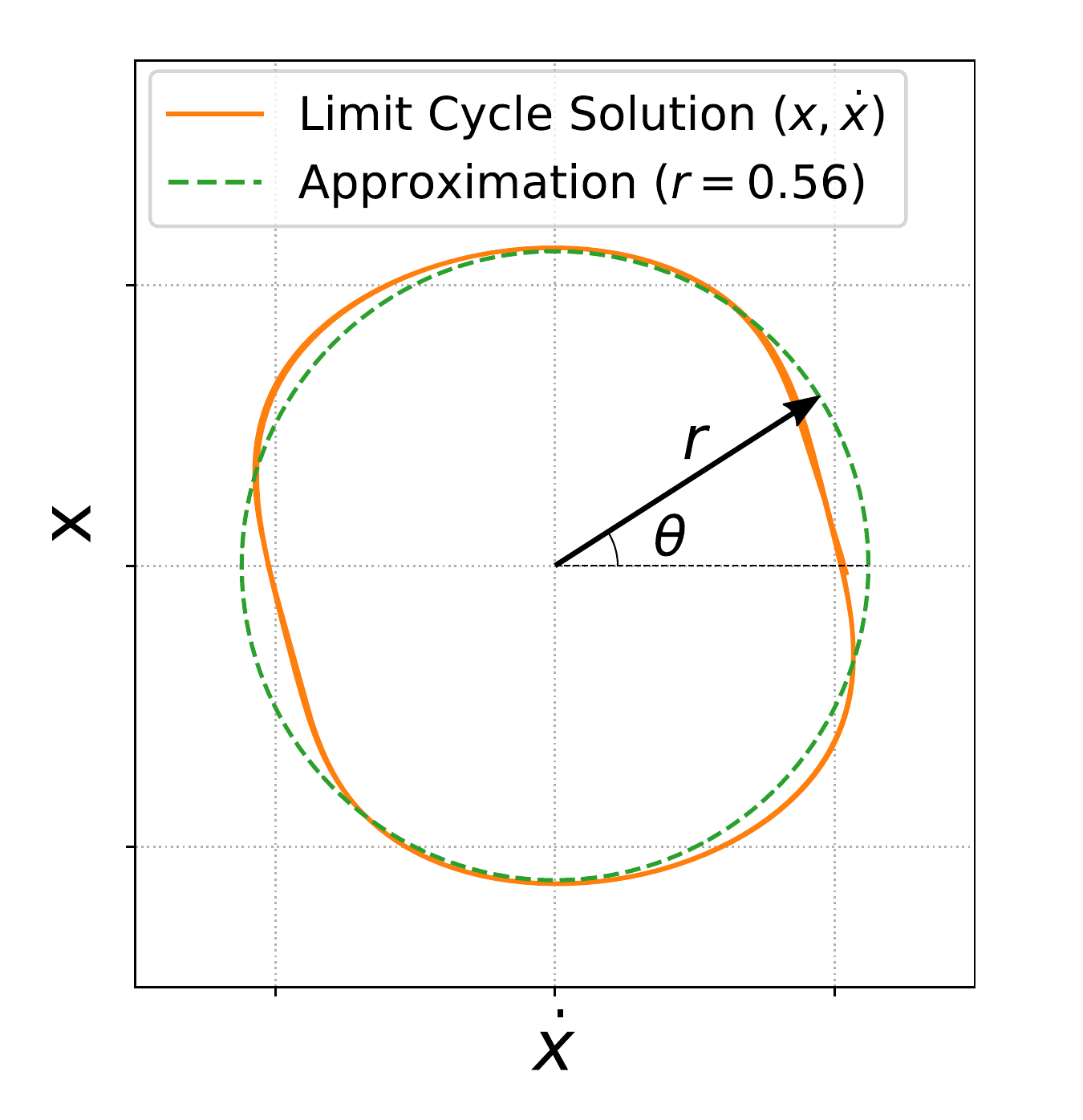}
	\caption{The limit cycle solution for the shape dynamics 
            $(x(t),\dot{x}(t))$ together with its limit cycle
            approximation (see~\eqref{eq:approxLimitCycle}).}
	\label{fig:limit_cycle}
\end{figure}

Consider the second-order equation \eqref{eq:i-Dynamics} for  the
shape variable $x$ under the open-loop periodic input $\tau(t)$ given
by~\eqref{eq:periodic_torque}. The following assumption is made
concerning its solution:

\begin{romannum}
    \item[\textbf{\textit{Assumption~A1}}] 
    Under periodic forcing $\tau(t)$ as in \eqref{eq:periodic_torque}, 
    the solution to \eqref{eq:i-Dynamics} is given by 
    an isolated asymptotically stable periodic orbit (limit cycle) 
    with period ${2\pi}/{\omega_0}$.
\end{romannum}
Denote the set of points on limit cycle as $\mathcal{P}\subset\mathbb{R}^{2}$. 
The limit cycle solution is parameterized by a phase coordinate 
$\theta\in[0,2\pi)$ in the sense that there exists an invertible map 
$X_{LC}:[0,2\pi) \to \mathcal{P}$ such that 
$X_{\text{LC}}(\theta(t))=(x(t),\dot{x}(t))$, 
where $\theta(t)=(\omega_0 t+\theta(0))$ mod $2\pi$ (see Figure~\ref{fig:limit_cycle}).  The definition
of the phase variable is extended locally in a small neighborhood of
the limit cycle by using the notion of isochrons~\cite{izhikevich2007dynamical}.



In terms of the phase variable, the first-order dynamics of the group variable~\eqref{eq:g-Dynamics} is expressed as
\begin{equation}
    \dot{q}(t) = \tilde f(X_{LC}(\theta(t)),u(t)) =: f(\theta(t),u(t))
    \label{eq:g-Dynamics_theta}
\end{equation}
and the observation model \eqref{eq:obsv} is expressed as
\begin{equation}
    \ud Z(t) = h(\theta(t)) \ud t + \sigma_W \ud W(t)
    \label{eq:obsv_theta}
\end{equation}
where 
$h(\theta) := \tilde h(X_{LC}(\theta))$. 

The optimal control problem \eqref{eq:opt-cont} 
in terms of the phase variable is given by
\begin{equation}
    J(\theta_0) 
    = \underset{u(\cdot)}{\min}\ {\sf E} \left[ 
    \int_0^\infty e^{-\gamma t} c(\theta(t),u(t)) \ud t \right]
    \label{eq:opt-cont-theta}
\end{equation}
where $c(\theta,u)=f(\theta,u)  + \frac{1}{2\epsilon} u^2$ and the minimum is 
over all control inputs $u(\cdot)$ adapted to the filtration
$\mathcal{Z}_t$.

In contrast to the original problem, the new problem is described by a
single phase variable.  With $u(t)\equiv 0$, the dynamics is
described by the oscillator model $\theta(t)=(\omega_0 t+\theta_0)$
mod $2\pi$.  Now, in the presence of (small) control input, the
dynamics need to be augmented by additional terms due to control:
\begin{equation}\label{eq:controlled_theta}
\ud \theta(t) = (\omega_0 + \epsilon g(\theta(t)),u(t)) \ud t
\end{equation}
 


\subsection{Step 2. Feedback particle filter (FPF)}
\label{subsec:FPF}

A feedback particle filter is constructed to obtain the posterior 
distribution of the phase variable $\theta(t)$ given the noisy observations
~\eqref{eq:obsv_theta}. The filter is comprised of $N$ stochastic processes 
$\{\theta^{i}(t):1\leq i\leq N\}$, where the value $\theta^{i}(t)\in[0,2\pi]$ 
is the state of the $i$-th particle (oscillator) at time $t$. 
The dynamics evolves according to
\begin{align}
    \ud \theta^i(t) & = \omega^i \ud t +  \epsilon g(\theta^i(t)),u(t))
    \ud t \nonumber
\\ & \quad + \frac{\v(\theta^i(t),t)}{\sigma_W^2} \circ 
    \left(\ud Z(t) - \frac{h(\theta^i(t)) + \hat{h}(t)}{2} \ud t \right)
    \label{eq:FPF}
\end{align}
where $\omega^i\sim \text{Unif}([\omega_0-\delta,\omega_0+\delta])$ is the 
frequency of the $i$-th oscillator, the initial condition 
$\theta^i(0)=\theta^i_0 \sim \text{Unif}([0,2\pi])$, 
and $\hat{h}(t) := {\sf E}[h(\theta^i(t))|\mathcal{Z}_t]$. 
The notation $\circ$ denotes Stratonovich integration.  
In the numerical implementation, $\hat{h}(t) \approx N^{-1} 
\sum_{i=1}^N h(\theta^i(t))$. 

Based on the FPF theory, the gain function $\v(\theta,t)$ is the solution of the Poisson equation, 
expressed in the weak-form as
\begin{equation}
    \Expect[\v (\theta(t),t) \psi'(\theta_t)|\clZ_t] 
    = \Expect[(h(\theta_t) - \hat{h}) \psi(\theta(t))|\clZ_t]
    \label{eq:Poisson-eq-weak-form}
\end{equation} 
for all test functions $\psi$ where $\psi'$ is the derivative. In the numerical implementation, 
the gain function is approximated by choosing $\psi$ from the Fourier basis 
functions $\{\sin(\theta),\cos(\theta)\}$ and approximating the expectations with empirical distribution 
of the oscillators~$\{\theta^{i}(t): 1\leq i\leq N\}$. 
The detailed numerical algorithm to approximate the gain function appears 
in 
~\cite{tilton2013multi}.

\medskip
\begin{remark}
There are two manners in which control input $u(t)$ affects the
dynamics of the filter state $\theta^i(t)$:
\begin{enumerate}
\item The $O(\epsilon)$ term $\epsilon g(\cdot,u(t))$ which models the effect
  of dynamics; 
\item The FPF update term which models the effect of sensor
  measurements.  This is because the control input $u(t)$ affects the
  state $x(t)$ (see~\eqref{eq:i-Dynamics}) which in turn affects the
  sensor measurements $Z(t)$ (see~\eqref{eq:obsv}).  
\end{enumerate}
\end{remark}


\subsection{Step 3. Q-learning }
\label{subsec:Q-learning}
%

Using the FPF, following the approach presented
in~\cite{mehta2013feedback}, express the partially observed optimal control problem~\eqref{eq:opt-cont-theta} 
as a fully observed optimal control problem in terms of oscillator
states $\thetaN(t)= (\theta^1(t),\ldots,\theta^N(t))$ according to
\begin{equation}
    \JN(\thetaN_0) 
    = \underset{u(\cdot)}{\min}\ {\sf E} \left[ 
    \int_0^\infty e^{-\gamma t} \cN(\thetaN(t),u(t)) \ud t \right]
    \label{eq:opt-cont-theta-i}\end{equation}
subject to~\eqref{eq:FPF}, where the cost $\cN(\thetaN,u) := \frac{1}{N}
\sum_{i=1}^N c(\theta^i,u)$ and the minimization is over all control laws 
adapted to the filtration $\mathcal{X}_t:=\{\theta^i(s); ~s\leq t, 
1\leq i\leq N\}$. The problem is now fully observed because the states of 
oscillators $\thetaN(t)$ are known. 


The analogue of the Q-function for continuous-time systems is 
the Hamiltonian function:
\begin{equation}\label{eq:Hamiltonian}
    \HN(\thetaN,u) = \cN(\thetaN,u) + \mathcal{D}_u \JN(\thetaN)
\end{equation}
where  $\mathcal{D}_u$ is the generator for~\eqref{eq:FPF} defined such that 
$\frac{\ud}{\ud t}\Expect[\JN(\thetaN(t))]=\mathcal{D}_u\JN(\thetaN(t))$.  


The dynamic programming principle for the discounted problem implies:
\begin{equation}\label{eq:dynamic-programming}
    \min_u~ \HN(\thetaN,u) = \gamma \JN(\thetaN)
\end{equation}
Substituting this into the definition of the Hamiltonian~\eqref{eq:Hamiltonian} 
yields the fixed-point equation:
\begin{equation}
    \mathcal{D}_u~\HminN(\thetaN) = \gamma (\cN(\thetaN,u) - \HN(\thetaN,u))
    \label{eq:fixed-pt}
\end{equation}
where $\HminN(\thetaN):=\min_u~\HN(\thetaN,u)$. 
This is the fixed-point equation that appears
in the Q-learning.

\medskip
\noindent
\textbf{Linear function approximation:}
Learning the exact Hamiltonian (Q-function) is an infinite-dimensional 
problem. Therefore, we set the goal to learn an approximation.  For
this purpose, consider a fixed set of $M$ real-valued functions 
$\{\phi^{(m)}(\theta,u)\}_{m=1}^M$.  The Hamiltonian is approximated as 
the linear combination of the basis functions as follows:
\begin{equation}
    \begin{aligned}
        \hatHN(\thetaN,u;w) & = \frac{1}{N} \sum_{i=1}^N w^T\phi(\theta^i,u) 
    \end{aligned}
    \label{eq:LFA}
\end{equation}
where $w\in \Re^M$ is a vector of parameters (or weights) and $\phi = (\phi^{(1)}, \phi^{(2)},\hdots, \phi^{(M)})^T$ is 
a vector of basis functions. 
Thus, the infinite-dimensional problem of learning the Hamiltonian function 
is converted into the problem of learning the $M$-dimensional weight
vector $w$. 

Define the point-wise Bellman error according to
\begin{equation}
    \begin{aligned}
        \mathcal{E}(\thetaN,u;w):=&\mathcal{D}_u\hatHminN(\thetaN;w) 
        \\&+ \gamma (\cN(\thetaN,u) - \hatHN(\thetaN,u;w))
    \end{aligned}
    \label{eq:Bellman-error}
\end{equation} 
where $\hatHminN(\thetaN;w):= \min_u~\hatHN(\thetaN,u;w)$. 



The following learning algorithm is proposed to find the optimal parameters
\begin{equation}
    \frac{\ud}{\ud t} w(t) 
    = -\frac{1}{2}\alpha(t) \nabla_w \mathcal{E}^2(\thetaN(t),u(t);w(t))
    \label{eq:update-w}
\end{equation}
where $\alpha(t)$ is the time-varying learning gain and $u(t)$ is chosen 
to explore the state-action space. For convergence analysis of the
Q-learning algorithm
see~\cite{tsitsiklis1999optimal,szepesvari1998asymptotic,even2003learning,moulines2011non}.

\medskip
Given a learned weight vector $w^*$, the learned optimal control policy is given by:
\begin{equation}\label{eq:learned-u}
\hat{u}^*(\thetaN;w^*) = \underset{v}{\arg\min}\ 
\hatHN(\thetaN,v;w^*) 
\end{equation}

\subsection{Information structure}

In order to implement the algorithm, one requires knowledge
of the following models:
\begin{enumerate}
\item A model for $g(\theta,u)$ which represents the reduced order
  model of dynamics as these affect the phase variable; 
\item A model for $h(\theta)$ which represents the reduced order model
  of the sensor.   
\end{enumerate}
These models are needed to implement the coupled oscillator
FPF~\eqref{eq:FPF}.   It is noted that both the models are described
with respect to the phase variable $\theta$.    

In the simulation results presented next, we assume
knowledge of $h(\theta)$ and ignore the term 
$\epsilon g(\theta^{i}(t),u(t))$ in the FPF.  The formal reason for ignoring the
term is that it is small ($O(\epsilon)$) compared to other terms, $O(1)$ frequency
$\omega_0$ and the update term.  Although, we assume knowledge of the
sensor model $h(\theta)$ for this paper, a learning algorithm could
also be implemented to learn the sensor model in an online manner.
This is the subject of future work.      

In numerical implementation, the generator $\mathcal{D}_u$ is approximated as 
\[
\mathcal{D}_u\hatHminN(\thetaN(t)) \approx \frac{\hatHminN(\thetaN(t+\Delta t)) 
- \hatHminN(\thetaN(t)) }{\Delta t}
\] 
where $\Delta t$ is the discrete time 
step-size in the numerical algorithm and $\{\theta^i(t)\}_{i=1}^N$ is the state of the oscillators at time $t
$. 

The reward function $f(\theta(t),u(t))$ in the cost function is approximated as
\[
f(\theta(t),u(t))=\dot{q}(t)\approx \frac{q(t+\Delta t) - q(t)}{\Delta
  t}
\] 
where $q(t)$ is available through the (black-box) simulator. 

The overall algorithm appears in Table~\ref{algm:QL-OptControl}.

\begin{algorithm}[t]
	\caption{Q-learning for Optimal Contol of Two-body System}
	\label{algm:QL-OptControl}
	\begin{algorithmic}[1]
		\Require Parameters in Table~\ref{tab:NumParam} and a simulator for~\eqref{eq:i-Dynamics}-\eqref{eq:g-Dynamics}-\eqref{eq:obsv}
		\Ensure Optimal control policy $\hat{u}^*(\thetaN;w)$.
		\State Initialize particles $\{\theta_0^i\}_{i=1}^N\sim
		\text{Unif}([0,2\pi])$;
		\State Initialize weight vector $w_0$ according to \eqref{eq:init-w}. 
		\For{$k=0$ to $t_T/\Delta t-1$}
		\State Choose control input $u_k$ according to \eqref{eq:u_sin};
		\State Input $u_k$ to the simulator and output $Z_k$ and $q_k$
		\State Update particles in FPF
		 \begin{equation*}
		\theta_{k+1}^i = \theta_k^i + \omega^i \Delta t 
		+ \frac{\v(\theta_k^i)}{\sigma_W^2}  
		(Z_{k+1} - Z_k - \frac{h(\theta_k^i)+\hat{h}_k}{2}
		\Delta t )
		\end{equation*}
		\State Compute the cost $c_k= \frac{1}{\Delta t} (q_{k+1}-q_k) + \frac{1}{2\epsilon}u_k^2$ 
		\State Compute \[\mathcal{D}_u \hatHminN_k=\frac{1}{\Delta t}
		(\hatHminN(\thetaN_{k+1};w_k)-
		\hatHminN(\thetaN_k;w_k))\]
		\State Compute Bellman error 
		\begin{equation*}
		\mathcal{E}_k=\mathcal{D}_u \hatHminN_k 
		+ \gamma(c_k - \hatHN(\thetaN_k,u_k;w_k))
		\end{equation*}
		\State Update weight vector 
		$w_{k+1}=w_k-\Delta t \alpha_k\mathcal{E}_k\nabla_w\mathcal{E}_k$
		\EndFor
		\State Output the learned control policy $\hat{u}^*(\thetaN;w_k)$ 
		from \eqref{eq:learned-u}.
	\end{algorithmic}
\end{algorithm}

\subsection{Approximate formula for optimal control input}
\label{subsec:analytical}
Assuming the knowledge  of the explicit form of the function $f(\theta,u)$, a semi-analytic approach is presented in~\cite{taghvaei2014coupled}
to derive the following approximate formula for optimal control:
\begin{equation}
    u^*(t) \approx \frac{\epsilon C}{N} \sum_{i=1}^N \cos(\theta^i(t))
    \label{eq:analytical-ctrl}
\end{equation}  
where $C$ is a constant that depends on the parameters of the model.
The formula was shown to be valid in the asymptotic  limit as 
$\epsilon \to 0$.  
The control policy \eqref{eq:analytical-ctrl} is implemented 
in~\cite{taghvaei2014coupled} where it is numerically shown to lead to clockwise rotation of the head body. 

The formula~\eqref{eq:analytical-ctrl} serves as a baseline for comparison with the results
of the numerical implementation of the Q-learning algorithm.

\addtolength{\textheight}{-1cm}   

\section{Numerics}
\label{sec:Numerics}

\begin{table}[tb]
	\centering
	\caption{Parameters for Numerical Simulation}
	\begin{tabular}{cccc}
		\hline
		\hline\noalign{\smallskip}
		Parameter & Description & \multicolumn{2}{c}{Numerical value} \\
		\hline
		\hline\noalign{\smallskip}
		\multicolumn{4}{c}{{\bf Two-body system}}\\
        $m_i$ & Mass of body $B_i$ &  $m_1=1$ & $m_2=\frac{1}{2}$\\\noalign{\smallskip}
        $I_i$ & Moment of inertia of body $B_i$ & $I_1=\frac{2}{3}$&$I_2=\frac{1}{6}$  \\\noalign{\smallskip}
		$d_i$ & Span of body $B_i$ & $d_1=1$&$d_2=1$  \\
		\hline\noalign{\smallskip}
        \multicolumn{4}{c}{Some auxiliary parameters}\\
        \noalign{\smallskip}
        \multicolumn{4}{c}{$\tilde{m}=\frac{m_1 m_2 }{m_1 + m_2} ,\quad\tilde{I}_i=I_i + \tilde{m} d_{i}^2
        ,\quad\lambda=\tilde{m} d_{1}d_{2}$}\\\noalign{\smallskip}
        \multicolumn{4}{c}{$A_{i}(x)=\tilde{I}_{i} + \lambda \cos(x)
        ,\quad \Delta(x)=\tilde{I}_{1}\tilde{I}_{2} - \lambda^{2}\cos^{2}(x)$}\\ 
        \hline
		\hline\noalign{\smallskip}
		\multicolumn{4}{c}{{\bf Dynamic model}}\\
        $\omega_0$ & Input torque frequency & \multicolumn{2}{c}{$1.0$} \\
        $\tau_0$ & Input torque amplitude & \multicolumn{2}{c}{$1.0$} \\
        $\kappa$ & Torsional spring coefficient & \multicolumn{2}{c}{$2.0$}\\
        $b$ & Viscous friction coefficient & \multicolumn{2}{c}{$0.1$}\\ 
        \hline
		\hline\noalign{\smallskip}
		\multicolumn{4}{c}{{\bf Sensor \& FPF}}\\
		$\Delta t$ & Discrete time step-size & \multicolumn{2}{c}{$0.01$} \\
        $\sigma_W$ & Noise process std. dev. & \multicolumn{2}{c}{$0.1$} \\
		$N$ & Number of particles & \multicolumn{2}{c}{$1000$} \\
		$\delta$ & Heterogeneous parameter & \multicolumn{2}{c}{$0.12$} \\
		\hline
		\hline\noalign{\smallskip}
		\multicolumn{4}{c}{{\bf Q-learning}}\\
		$t_T$ & simulation terminal time & \multicolumn{2}{c}{$100\text{·}2\pi\text{/}\omega_0$} \\
		$\gamma$ & Discount rate & \multicolumn{2}{c}{$1.0$} \\
		$\epsilon$ & Control penalty parameter & \multicolumn{2}{c}{$1.0$} \\
		$\alpha_k$ & Learning gain & \multicolumn{2}{c}{$0.5$} \\
		$A$ & Control exploration amplitude & \multicolumn{2}{c}{$0.25$} \\
        \hline 
	\end{tabular}
	\label{tab:NumParam}
\end{table}

In this section, we present the numerical results.  These results
illustrate the i) phase modeling; ii) the performance of coupled
oscillator FPF; and iii) the performance of Q-learning algorithm. 
The numerical results are based on the use of
Algorithm~\ref{algm:QL-OptControl}.  The simulation parameters are
tabulated in Table~\ref{tab:NumParam}.

\subsection{Simulator}
\begin{figure*}[t]
	\centering
	\begin{tabular}{cc}
		\subfigure[]{
			\includegraphics[width=1.0\columnwidth,keepaspectratio=true]
			{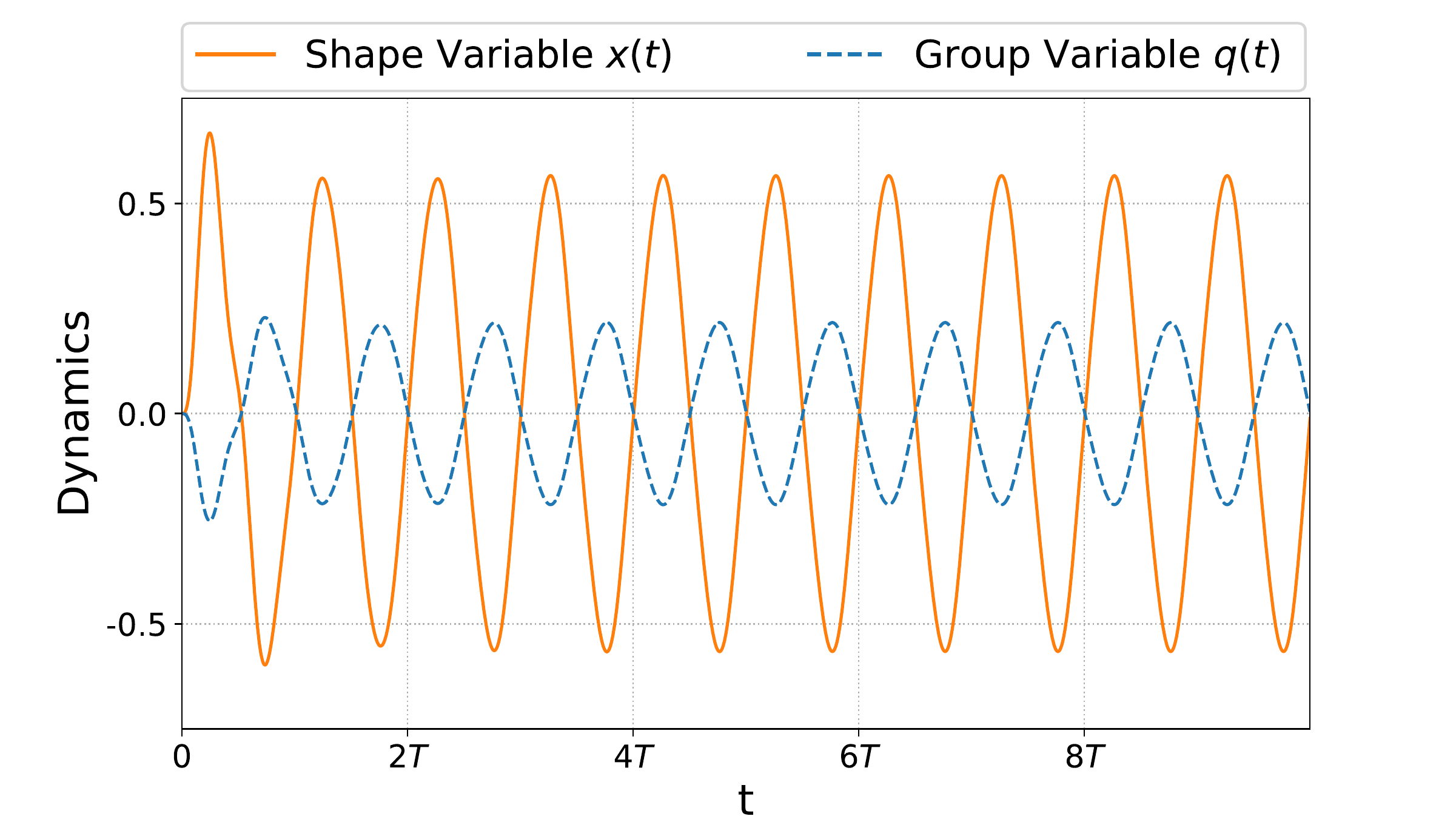}
			\label{fig:dynamics}
		}
		\subfigure[]{
			\includegraphics[width=1.0\columnwidth,keepaspectratio=true]
			{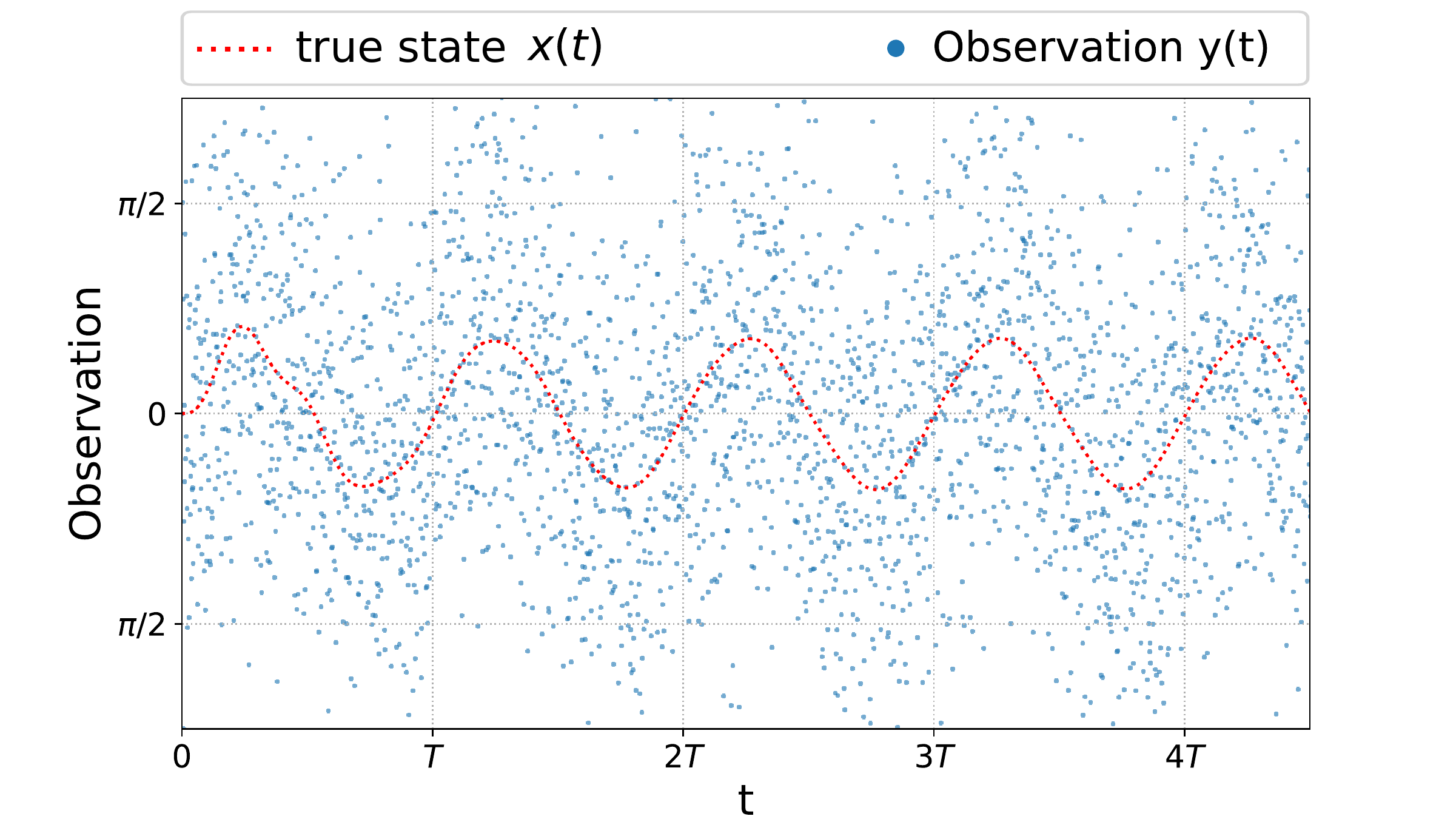}
			\label{fig:obsv}
		}
	\end{tabular}
    \caption{Summary of numerical results for the two-body system simulator: 
            (a) Trajectory for the shape variable 
            $x(t)$ and the the head orientation 
            $q(t)$ under no control input ($u(t)\equiv 0$); 
            (b) Observation process $y(t)$.}
	\label{fig:simulator}
\end{figure*}

The simulator takes the control input $u(t)$ and outputs the shape 
variable $x(t)$, the head orientation $q(t)$, and the observation 
$Z(t)$ according to~\eqref{eq:i-Dynamics}, \eqref{eq:g-Dynamics}, 
and~\eqref{eq:obsv} respectively. 
The explicit form of the simulated dynamics~\eqref{eq:i-Dynamics}-\eqref{eq:g-Dynamics} are as follows:
	\begin{align*}
	    \ddot{x}&=\frac{1}{\Delta}\left[-\lambda \sin(x)\frac{A_{1}A_{2}}{A_1
		+A_2}\dot{x}^{2}+(A_1+A_2)(\tau(t)- \kappa x -b\dot{x})\right]
	\\
	    \dot{q} &= \frac{\tilde{I}_2 +
		\lambda\cos(x)}{\tilde{I}_1+\tilde{I}_2+2\lambda
		\cos(x)}\dot{x}
	\end{align*}
where the parameters are defined in Table~\ref{tab:NumParam};
cf.,~\cite{taghvaei2014coupled} for additional details on modeling.  
The resulting state trajectory $x(t)$  and head orientation 
$q(t)$, under periodic open-loop torque $\tau(t)$
(see~\eqref{eq:periodic_torque}), 
are depicted in Figure~\ref{fig:dynamics}. 
It is observed that without control input, the head orientation oscillates, without any net change.

The observation signal $y(t):=(Z(t+\Delta t)-Z(t))/\Delta t$, with the step-size $\Delta t=0.01$ is depicted in Figure~\ref{fig:obsv}.
The observation model is taken as $\tilde{h}(x,\dot x)=x$.  The noise strength 
$\sigma_w = 0.1$.


\subsection{Phase modeling}
\label{subsec:Num-PM-FPF}
The limit cycle solution for the shape dynamics~\eqref{eq:i-Dynamics}  is depicted in Figure~\ref{fig:limit_cycle}. 
%
%
The map $\tilde{X}_{LC}(\theta(t)) = (x(t),\dot{x}(t)) $  is approximated as
\begin{align}
\tilde{X}_{LC}(\theta) & \approx (r\sin(\theta),r \omega_0 \cos(\theta))
\label{eq:approxLimitCycle}
\end{align}
where $r=0.56$ is numerically determined.

\subsection{Coupled oscillator feedback particle filter} 
\label{subsec:Num-FPF}


 
\begin{figure*}[t]
	\centering
	\begin{tabular}{ccc}
		\subfigure[]{
			\includegraphics[width=1.1\columnwidth,keepaspectratio=true]
			{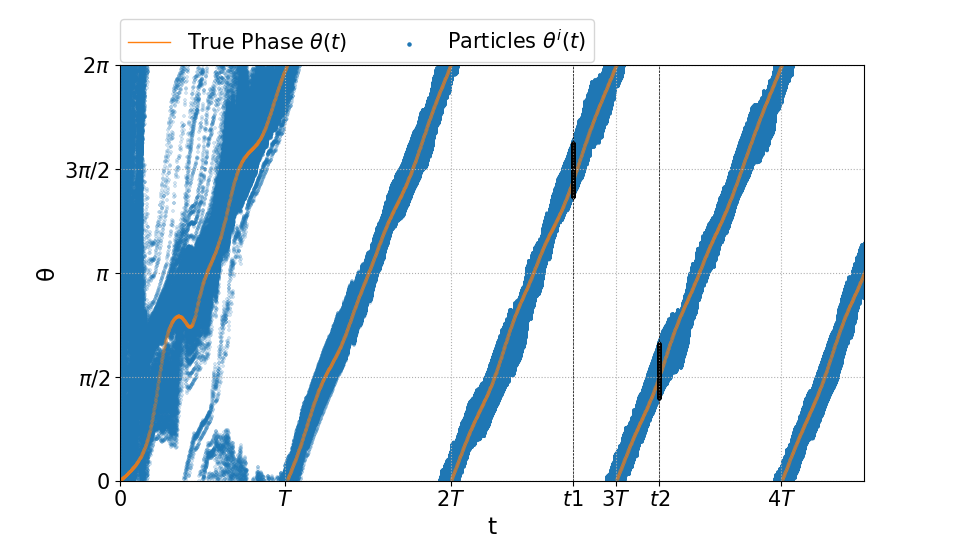}
			\label{fig:FPF}
		}
		\subfigure[]{
			\includegraphics[width=0.8\columnwidth,keepaspectratio=true]
			{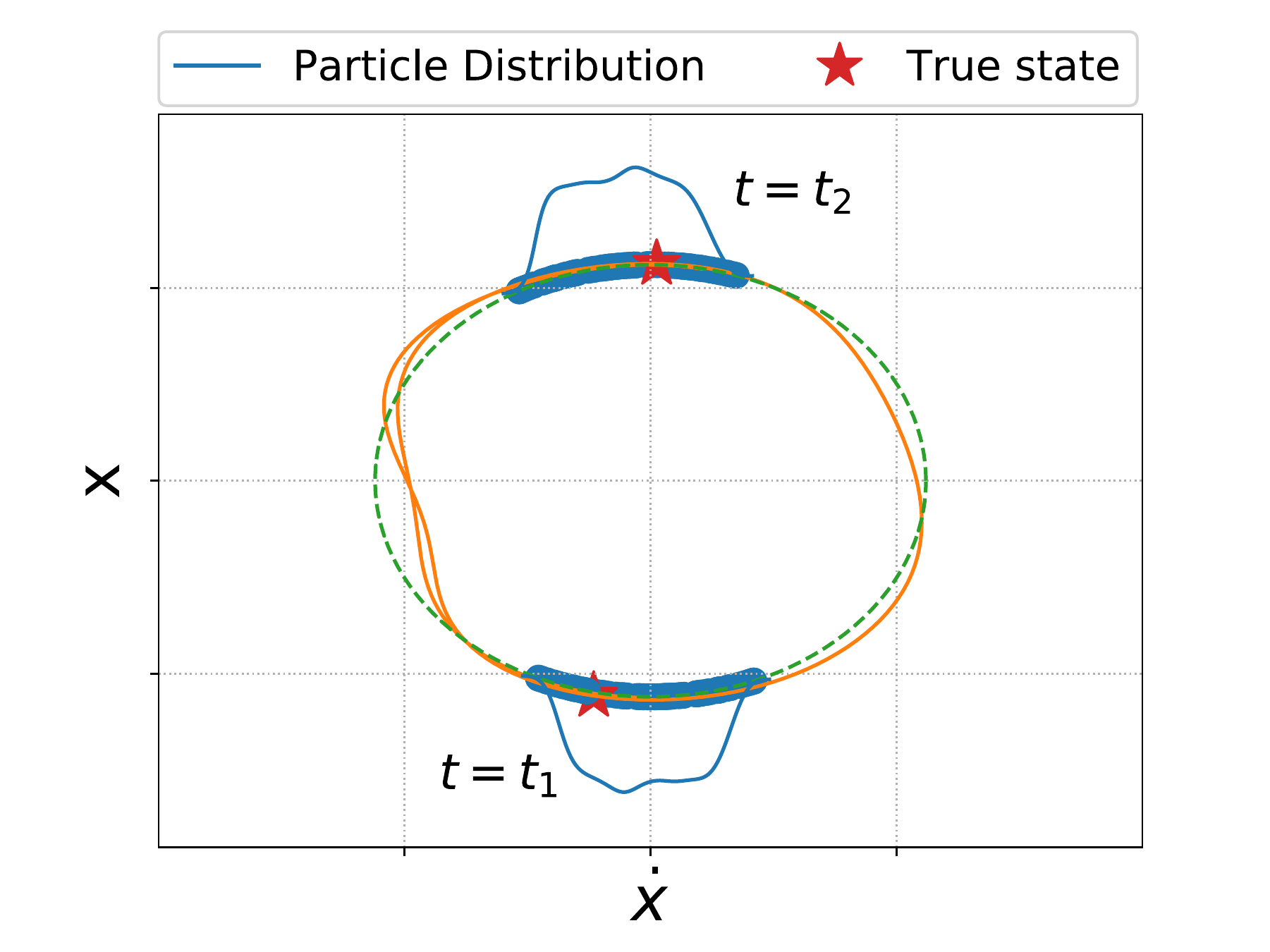}
			\label{fig:LimCycle_particles}
		}
	\end{tabular}
	\caption{Summary of estimation results:
            (a) Time trace of $N=1000$ particles in the FPF algorithm compared with the true 
            phase $\theta(t)$; 
            (b) Empirical distribution of
            the particles compared with the true state $(x,\dot x)$  at two time instants $t_1$ and $t_2$.  
        }
	\label{fig:LimCycle & FPF}
\end{figure*}

%
The trajectories of $N=1000$ particles in the FPF algorithm~\eqref{eq:FPF} are depicted in Figure~\ref{fig:FPF}. 
The initial conditions $\theta_0^i$ 
are drawn from a uniform distribution in $[0,2\pi]$.
The initial transients due to the initial condition converge rapidly.
The true phase variable $\theta(t)$ is also depicted
as a solid line. It is observed that the ensemble of particles 
synchronize and track the true phase.



The true state $(x(t),\dot{x}(t))$, along with the empirical distribution of the particles, at two time instants, are shown in Figure~\ref{fig:LimCycle_particles}. The particles are positioned on the approximate limit cycle 
map~\eqref{eq:approxLimitCycle}. 
These results show that the filter is able to track the 
true state on the limit cycle accurately.

\subsection{Q-learning}
\label{subsec:Num-QL}
To approximate the Hamiltonian function in~\eqref{eq:LFA}, the basis
functions are selected to be the product of Fourier basis functions of
$\theta$ and polynomial functions of $u$  as follows:
\begin{align*}
\phi(\theta,u)= \big( &\cos(\theta), \sin(\theta), \cos(2\theta), \sin(2\theta), \\
&u\cos(\theta), u\sin(\theta), u\cos(2\theta), u\sin(2\theta), 
\frac{1}{2}u^2\big)^T
\end{align*}
The weight vector $w_0=\Big(w_0^{(1)}, ..., w_0^{(9)}\Big)^T$ is initialized 
randomly as follows:
\begin{equation}
\begin{aligned}
w_0^{(m)}&\sim\text{Unif}([-1,1]) \quad \text{for } m=1,\ldots, 8\\
w_0^{(9)}&\sim\text{Unif}([0.9,1.1])
\end{aligned}
\label{eq:init-w}
\end{equation}
The reason for choosing the particular initialization for
$w_0^{(9)}$is to avoid numerical issues due to large control values
whenever $w_0^{(9)}$ is small. 

The exploration control input $u(t)$ to be used in~\eqref{eq:update-w} is chosen as a 
combination of sinusoidal functions with irrationally related frequencies 
as follows:
\begin{equation}
u(t) = A\sin(\omega_0t) + A\sin(\pi\omega_0t)  
\label{eq:u_sin}
\end{equation}
The rationale for doing so is to explore the state-action space~\cite{bertsekas1996neuro}. 

\begin{figure*}[t]
	\centering
	\begin{tabular}{ccc}
		\subfigure[]{
			\includegraphics[width=1.0\columnwidth,keepaspectratio=true]
			{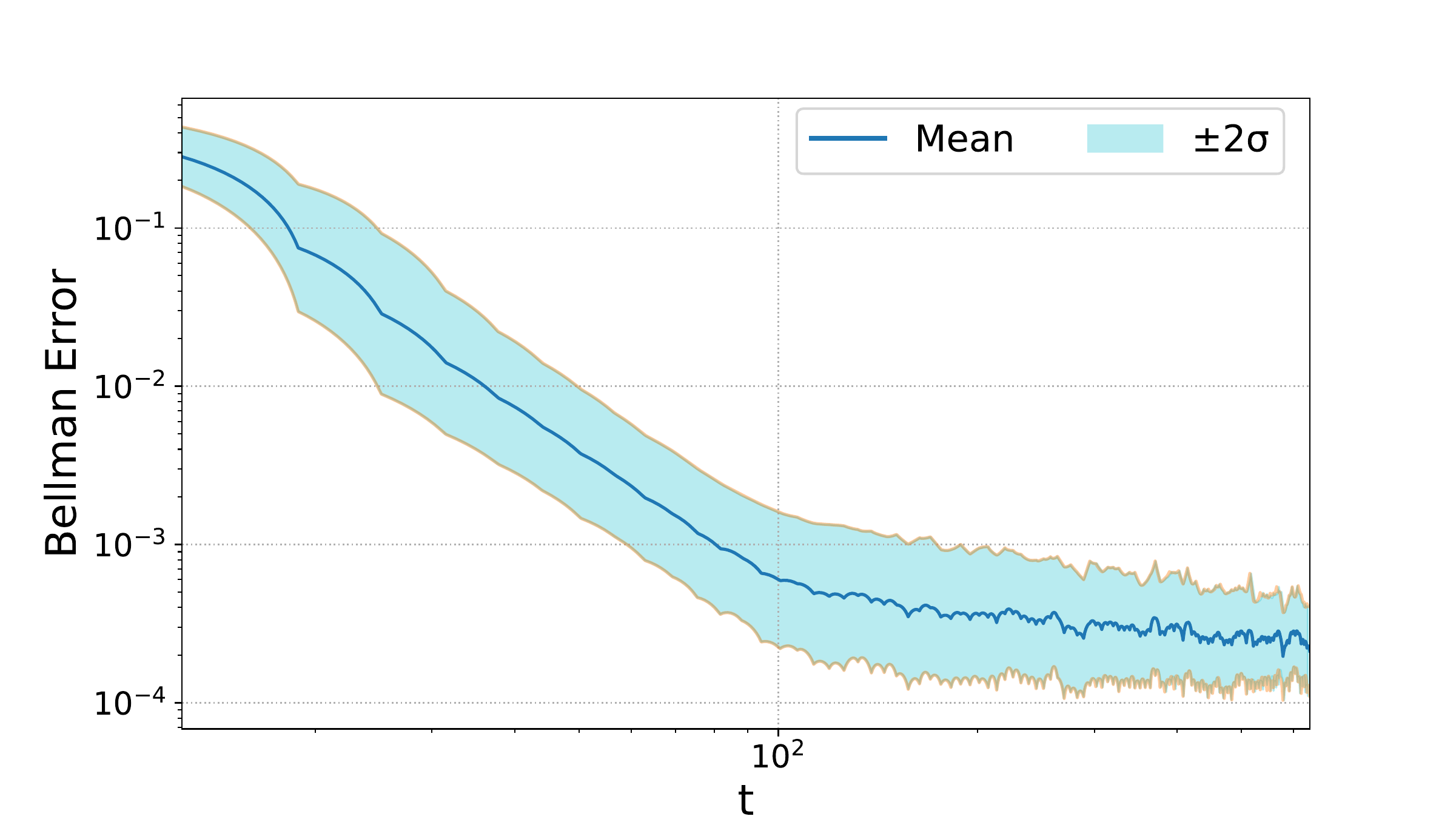}
			\label{fig:BE}
		}
		\subfigure[]{
			\includegraphics[width=1.0\columnwidth,keepaspectratio=true]
			{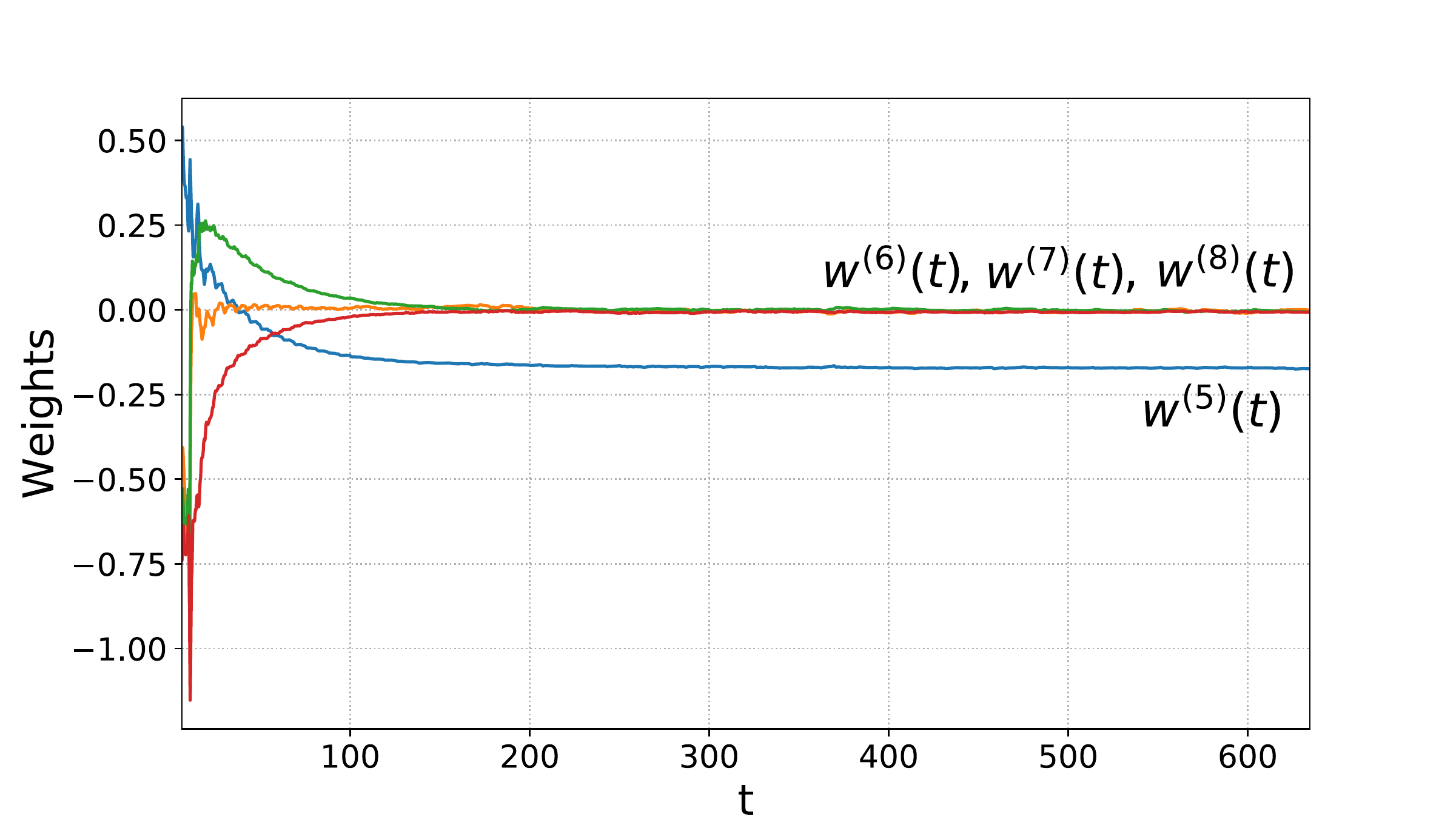}
			\label{fig:weight_vector}
		}
	\end{tabular}
	\caption{Summary of Q-learning results:
            (a) The Bellman error as a function of time; 
            (b) Convergence of the four weight components in the learned optimal control law~\eqref{eq:learning-ctrl}.
           }
	\label{fig:convergence}
\end{figure*}

\begin{figure*}[t]
	\centering
	\begin{tabular}{cc}
		\subfigure[]{
			\includegraphics[width=1.0\columnwidth,keepaspectratio=true]
			{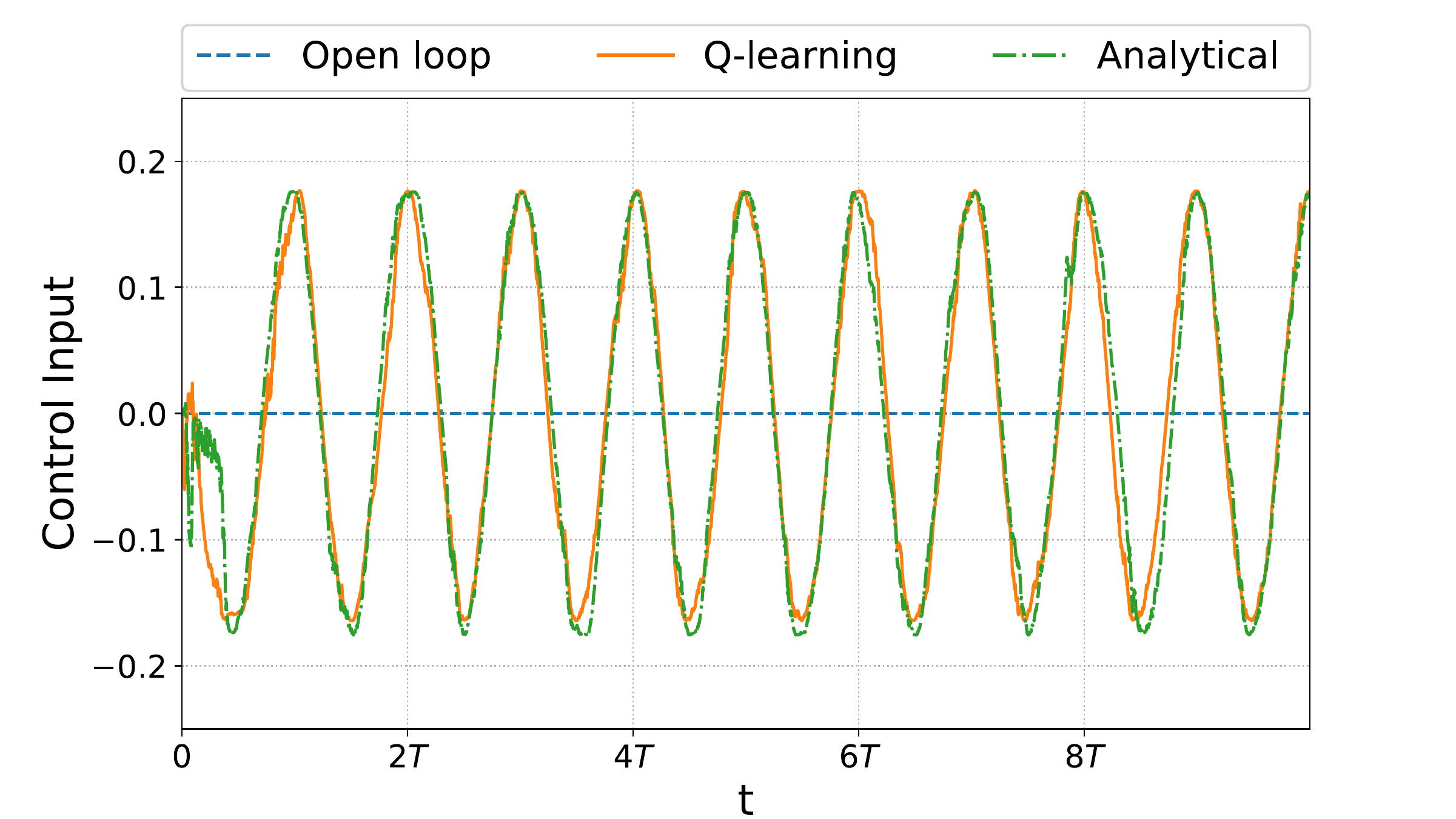}
			\label{fig:rand.u}
		}
		\subfigure[]{
			\includegraphics[width=1.0\columnwidth,keepaspectratio=true]
			{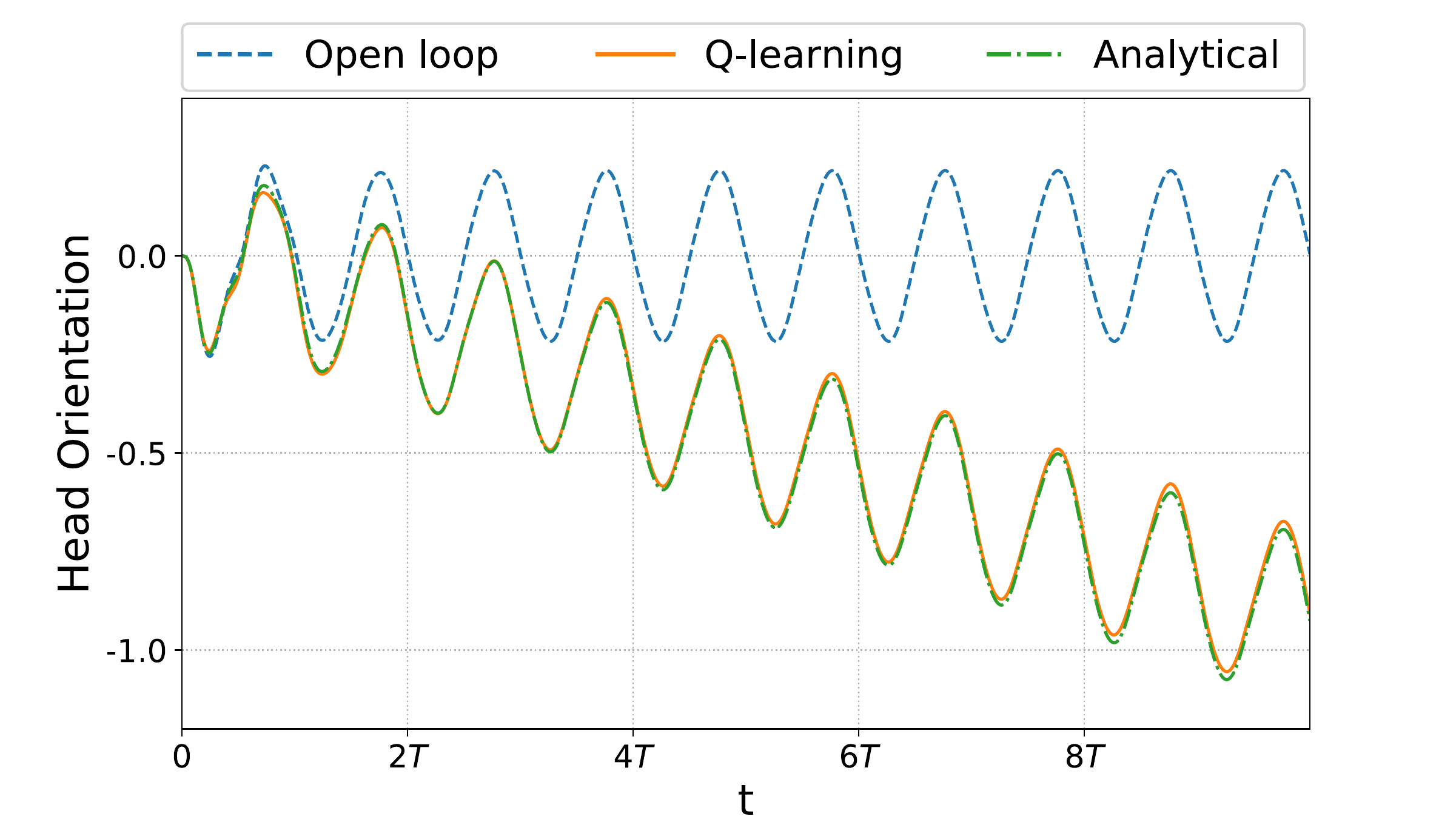}
			\label{fig:rand.q}
		}
	\end{tabular}
	\caption{Summary of control results: 
        (a) Comparison of the control input learned from Q-learning 
        \eqref{eq:learning-ctrl} and the semi-analytical optimal control input~\eqref{eq:analytical-ctrl}; 
        (b) Time trace of the orientation with no control input ($u(t)\equiv 0$), semi-analytical control input, and the learned control input.}
	\label{fig:u & q}
\end{figure*}

The $L^2$-norm of the point-wise Bellman error~\eqref{eq:Bellman-error}, over the $j$-th period is denoted as $e_j$ and defined to be:
\begin{equation}
e_j := 
 \frac{1}{T}
\int_{(j-1)T}^{jT}\Big|\mathcal{E}(\thetaN(t),u(t);w(t))\Big|^2 \ud t
\label{eq:BE-mean}
\end{equation}
The average of the error $e_j$ and its variance over fifty Monte-Carlo runs are depicted in Figure~\ref{fig:BE}.  
It is observed that the Bellman error drops by over three 
orders of magnitude.  This suggests that the algorithm is able to
learn the Hamiltonian function that solves the dynamic programming
fixed-point equation~\eqref{eq:fixed-pt}.



The learned optimal control policy~\eqref{eq:learned-u} in terms of the selected basis functions is given by: 
\begin{equation}
\begin{aligned}
    \hat{u}^*(\thetaN;w)  = &-\frac{1}{N}\sum_{i=1}^N \left(
    \frac{w^{(5)}}{w^{(9)}}\cos(\theta^i) + \frac{w^{(6)}}{w^{(9)}}\sin(\theta^i) \right)\\
     &-\frac{1}{N}\sum_{i=1}^N \left(   \frac{w^{(7)}}{w^{(9)}}\cos(2\theta^i) + \frac{w^{(8)}}{w^{(9)}}\sin(2\theta^i)\right)
    \label{eq:learning-ctrl}
\end{aligned}
\end{equation} 
The traces of the four components of the weight vector
$\{w^{(5)}(t),w^{(6)}(t),w^{(7)}(t),w^{(8)}(t)\}$ 
are depicted in Figure~\ref{fig:weight_vector}. 
It is observed that the components (weights) converge quickly. Also, the component
$w^{(5)}(t)$, which corresponds to
$\cos(\cdot)$, dominates while other components converge to zero.

Figure~\ref{fig:rand.u} depicts the learned optimal control policy as
a function of time.  Also depicted is a comparison  with the
semi-analytical solution~\eqref{eq:analytical-ctrl}. 
It is observed that the learned optimal control coincides with the
analytical formula in terms of both phase and amplitude.

Figure~\ref{fig:rand.q} depicts the resulting head orientation with i)
the learned optimal control policy~\eqref{eq:learning-ctrl}; ii) the
analytical control law~\eqref{eq:analytical-ctrl}; and iii) with no control input ($u(t)\equiv 0$).
It is observed that the learned optimal control input induces
nearly the same net change in the head orientation as the analytical
control law. That is, the Q-learning algorithm is able to learn the
optimal control policy to rotate the head body clockwise.




.
\section{Conclusions and Future Work}
\label{sec:Conclusions}

\label{subsec:Concl}

We introduced a coupled oscillator-based framework for learning the
optimal control of periodic locomotory gaits.  The framework does not
require knowledge of the explicit form of the dynamic models or
observation of the full state.  The framework was illustrated on the
problem of turning the two-body planar system. 
One direction for future work, is to apply the framework to more
complicated models such as coupled kinematic chains and continuum
models.  Another direction of future work is to consider more advanced
tasks such as turning to a certain angle or locating a target.

\bibliographystyle{plain}
\bibliography{reference}

\end{document}